\documentclass[10pt,conference]{IEEEtran}

\usepackage{cite}
\usepackage{amsmath,amssymb,amsfonts}
\usepackage{algorithmic}
\usepackage{graphicx}
\usepackage{textcomp}
\usepackage{xcolor}
\usepackage{listings}
\usepackage{xcolor}
\usepackage{xspace}
\usepackage[inline,shortlabels]{enumitem}
\usepackage{multirow}
\usepackage{calc}
\usepackage{hyperref}
\usepackage{booktabs}
\usepackage{lscape}
\usepackage{comment}
\usepackage{cleveref}
\usepackage{tabularx}
\usepackage{balance}
\usepackage{tcolorbox}
\usepackage[symbol]{footmisc}
\usepackage{orcidlink}

\tcbset{size=small,top=1mm,bottom=1mm}

\def\BibTeX{{\rm B\kern-.05em{\sc i\kern-.025em b}\kern-.08em
    T\kern-.1667em\lower.7ex\hbox{E}\kern-.125emX}}

\newcommand{\ie}{\emph{i.e.\@}\xspace}
\newcommand{\eg}{\emph{e.g.\@}\xspace}

\newcommand{\mypara}[1]{\paragraph{#1}}
\def\paragraph#1{\smallskip\noindent\textbf{\textit{#1.}}\xspace}

\newcommand{\red}[1]{\textcolor{red}{#1}}

\newcommand{\rh}[1]{\textcolor{orange}{[RH: #1]}}

\newcommand{\comp}[0]{Comp\xspace}
\newcommand{\bio}[0]{Bio\xspace}
\newcommand{\earth}[0]{Earth\xspace}
\newcommand{\soc}[0]{Soc\xspace}

\usepackage{ifthen}
\usepackage[normalem]{ulem} 

\newboolean{showedits}
\setboolean{showedits}{false} 
\ifthenelse{\boolean{showedits}}
{
	\newcommand{\ins}[1]{\textcolor{blue}{\uline{#1}}}
    \newcommand{\inscite}[1]{\textcolor{blue}{#1}}
	\newcommand{\del}[1]{\textcolor{red}{\sout{#1}}} 
	\newcommand{\chg}[2]{\textcolor{red}{\sout{#1}}{\ra}\textcolor{blue}{\uline{#2}}} 
}{
	\newcommand{\ins}[1]{#1} 
    \newcommand{\inscite}[1]{#1} 
	\newcommand{\del}[1]{} 
	\newcommand{\chg}[2]{#2}
}
    
\begin{document}

\title{How Scientists Use Jupyter Notebooks: Goals, Quality Attributes, and Opportunities}

\author{
\IEEEauthorblockN{
Ruanqianqian (Lisa) Huang\IEEEauthorrefmark{1}\orcidlink{0000-0002-4242-419X}, 
Savitha Ravi\IEEEauthorrefmark{1}\orcidlink{0009-0006-9756-8832}, 
Michael He, 
Boyu Tian, 
Sorin Lerner\orcidlink{0000-0003-3957-0628}, 
Michael Coblenz\orcidlink{0000-0002-9369-4069}
}
\IEEEauthorblockA{
\textit{University of California San Diego}\\
La Jolla, CA, United States \\
\{r6huang, s2ravi, mih024, btian, lerner, mcoblenz\}@ucsd.edu}
}


\maketitle

\begin{abstract}

Computational notebooks are intended to prioritize the needs of scientists, but little is known about how scientists interact with notebooks, what requirements drive scientists' software development processes, or what tactics scientists use to meet their requirements. We conducted an observational study of 20 scientists using Jupyter notebooks for their day-to-day tasks, finding that scientists prioritize different quality attributes depending on their goals. A qualitative analysis of their usage shows (1) a collection of goals scientists pursue with Jupyter notebooks, (2) a set of quality attributes that scientists value when they write software, and (3) tactics that scientists leverage to promote quality. In addition, we identify ways scientists incorporated AI tools into their notebook work. From our observations, we derive design recommendations for improving computational notebooks and future programming systems for scientists. Key opportunities pertain to helping scientists create and manage state, dependencies, and abstractions in their software, enabling more effective reuse of clearly-defined components.

\end{abstract}


\footnotetext[1]{Co-first authors.}

\begin{sloppypar}
\section{Introduction}

Unlike traditional development environments, computational notebooks interleave program text with program output in a linear flow of content. Notebooks divide programs into \emph{cells}, enabling users to execute cells and see output in any order they choose. Computational notebooks, such as Jupyter~\cite{Jupyter}, have enabled thousands of users to create millions of notebooks~\cite{Pimentel2019:LargeScale, granger2021jupyter} to explore and communicate ideas in a form interleaving code, output, prose, and sometimes multimedia. These tools facilitate \emph{end-user software engineering}~\cite{Burnett2010:End,BARRICELLI2019101}: creation of critical software systems by people whose main focus is completing tasks, not authoring software.


Despite the broad adoption of computational notebooks, they only provide a thin layer over existing runtime environments. Code in cells can side-effect the environment, impeding reproducibility when users execute cells out of order. Cells provide no abstractions; they are not invokable, for example, and their code shares a scope with surrounding code. Version control is rarely used~\cite{kery2017variolite,Pertseva2024:Theory}, even by notebook users who use version control in their other work. Previous research~\cite{Lau2020:Design} and commercial implementation~\cite{Observable} exploring alternative designs have not yet produced popular competitors.

Why do computational notebooks still serve the needs of their users better than existing alternatives? 
We hypothesize that this is not merely a problem of challenges transitioning from popular, well-documented tools; instead, we do not yet understand enough about \emph{how} computational notebooks meet their users' needs. To lay a foundation for a new generation of tools, we seek to understand \ins{notebook users' \emph{non-functional requirements}---}\emph{properties that notebook users value}\del{,}\ins{ and} \emph{how they promote the properties they value}\del{, }\ins{---}and \emph{design opportunities} that could lead to more effective \ins{notebook }systems.\del{ We frame this in terms of \emph{non-functional requirements}: what properties of programs do notebook users value, and what tactics do they use to promote those properties? }

\ins{This paper focuses on \emph{scientific users} of computational notebooks, \ie, those who write scientific programming code in notebooks. 
Prior work suggests that scientists struggle with keeping effective work in notebooks and transitioning from notebooks to other general-purpose programming tools; nonetheless, notebooks are widely popular among scientists}~\cite{Pertseva2024:Theory}\ins{.}
To understand how and why computational notebooks benefit \del{their users}\ins{scientists}---and identify opportunities for improvement---we focus on \del{two}\ins{three} research questions:


\begin{description}
    \item [\ins{RQ1}] \ins{What goals do scientists pursue in notebooks?}
    \item [RQ\del{1}\ins{2}] What non-functional requirements do scientific users of computational notebooks prioritize?
    \item [RQ\del{2}\ins{3}] What tactics do scientists use to create notebooks that meet their quality requirements while also facilitating their scientific discovery process?
\end{description}

Past \del{studies}\ins{investigations} have \del{surveyed and interviewed}\ins{studied} users of Jupyter notebooks\ins{, focusing broadly on general data workers}~\cite{DeSantana2024:Bug,Chattopadhyay2020:Whats,kery2017variolite}. However, like Ko et al. in their study of Java programming~\cite{ko2006exploratory}, we wanted to understand the interaction-level techniques that \del{programmers}\ins{scientists} use to create software that meets their requirements. Therefore, we conducted an observational study of how individual scientists use Jupyter notebooks to accomplish realistic day-to-day tasks. In our IRB-approved study, we recruited 20 participants to let us watch them do their work. To understand how software-related training affects use, we recruited both computer scientists and non-computer scientists.  


%
We first observed how each participant worked with Jupyter notebooks using their own setup and workflow for up to 45 minutes. Then, because any limited observation period could have resulted in unfinished work, we asked them to wrap up the work for future continuation. Finally, we interviewed them about their experience with Jupyter notebooks during the study as well as overall perceptions.

Leveraging techniques from constructivist grounded theory~\cite{Charmaz2014:Constructing}, we noted memos of each observation and coded every notebook-related action as well as every notebook-related comment made by participants. In \autoref{sec:results}, we present eight quality attributes that our participants valued: clarity, explorability, reusability, reproducibility, correctness, performance, debuggability, and collaboration. 
We describe tactics, such as implementing each task in one cell, that participants used to promote quality. 
Additionally, we show how participants incorporated AI tools into notebook work in scientific settings.
Finally, we discuss design opportunities that are revealed by this framing of computational notebook use. 



%

Cell structure and out-of-order execution distinguish notebooks from other programming environments and promote explorability, particularly for code that can be slow to run. Unfortunately, they simultaneously inhibit aspects of correctness and reproducibility, which scientists also value. However, future notebook systems could empower users to choose their tradeoffs more precisely by controlling scope, dependencies, and in what cases cells are evaluated automatically.
%

While our work may not render a complete description of how people use Jupyter notebooks in every possible way, our findings are contextualized in realistic tasks and can inspire future improvements to Jupyter notebooks and similar end-user programming environments.

In summary, this paper's key contributions are:
\begin{itemize}
    \item Eight quality attributes that scientists value when creating and maintaining  notebooks from an an observational study of 20 participants performing their own tasks\ins{;}
	\item Tactics that scientists use to promote software quality in Jupyter notebooks\ins{;}
	\item Opportunities for improvement in notebook tools that could promote quality in software written by scientists\ins{.}
\end{itemize}
\section{Methods}


\subsection{Participants}

%
%
We recruited participants from various disciplines via institutional mailing lists, messaging platforms, and snowball sampling.
Since we aimed to study users with different domain expertise, programming experience, and software engineering experience, we did not impose screening constraints other than requiring (1) prior experience with Jupyter notebooks and (2) the ability to demonstrate a realistic task to be done in Jupyter notebooks during the study.
We explicitly hoped to recruit people with varying experience in software engineering, so we recruited half of the participants from the computer science discipline and the other half from other scientific disciplines. We stopped recruiting participants after we had achieved saturation\ins{, following standard practice}\inscite{~\cite{Charmaz2014:Constructing}}: \ins{After each study, we compared field notes to existing observations to construct theoretical categories and their properties; the last four participants did not lead to new categories or properties.}\del{the most recent participants did not reveal any new quality attributes or actions.}


\subsection{Procedure}

Each study session included two parts: a 60-minute observation and a 30-minute interview. 
We conducted the studies in person and over Zoom, recording participants' screens for analysis.
Two authors ran the studies and took turns to lead. Participants received a \$23 gift card after the study.

During the observational portion, participants worked on tasks of their choosing (more in \autoref{subsec:task}) using Jupyter notebooks\del{, and the remaining time was spent for the interview}. We prepared a backup task in case participants did not have a task ready, but we never used it.
We asked participants to explain their tasks as they worked, in the style of think-aloud. However, past observations of programmers have shown that additional depth is needed~\cite{Coblenz2021:PLIERS}, so while working on the task, the experimenter asked questions as needed about the subject's workflow when using notebooks and various aspects of the task. 
After about 45 minutes of observation, we asked the participant to start wrapping up their work to continue later. Most tasks can take many work sessions to complete, and \del{work }sessions can be hours long. To observe how users initiated the cleanup process in the limited time span of a study, we asked participants to spend up to 15 minutes wrapping up their work so they could easily pick up from where they left off; \ins{the 15-minute threshold was determined via pilot studies, and no study participants needed all 15 minutes. In the cases where we observed no wrap-up activities, we asked about participants' usual practices.} 

We conducted a semi-structured interview afterward.
Our questions pertained to five topics: notebook reproducibility, 
notebook understandability,
the ability of the participant or collaborators to continue working on the notebook, 
the expected longevity of the current notebook, 
and changes in workflow when using computational notebooks compared to scripts. 
%

\subsection{Protocol Development Process}\label{subsec:protocol-development}
%
%
We started the study with one very open-ended question: \emph{How do people use Jupyter notebooks for their day-to-day tasks?} \ins{Prior work on notebook use and pain points}~\cite{Chattopadhyay2020:Whats, Rule2018:Exploration}\ins{ motivated our initial interview topics. }We used a qualitative analysis method with key techniques from constructivist grounded theory~\cite{Charmaz2014:Constructing}, including open coding, theoretical coding, and memo writing. Following the \emph{theoretical sampling} approach, we recruited participants in areas in which we lacked data, and we refined the questions we asked participants \ins{per Charmaz's guidelines}~\cite{Charmaz2014:Constructing}\ins{ based on emerging patterns }to enable a deeper understanding of topics for which we had not yet gathered sufficient data\del{.}\ins{; while this led to new questions asked in later studies, our IRB advised us that new questions within the subject matter of the study did not need further review.} 

After ten sessions, we analyzed the memos that we had written and found that participants had varying attitudes towards using Jupyter notebooks: some used ad hoc organizational techniques, caring little about readability and/or reproducibility; others attended to organizing the content inside a notebook, ensuring attributes such as modularity, readability, and reproducibility.
For example, P1 mentioned \emph{``notebooks have no longevity;''} do all notebook users hold the same low expectations for the longevity of notebooks? Expectations about longevity could have implications on how users prioritize readability and reproducibility.
With these questions in mind, starting the 11th session, we added additional questions to our protocol asking participants about their expectations for the notebook's longevity.
We also observed that most participants conducted cleanup tasks such as adding Markdown headings, leaving notes on to-dos both inside and outside the notebook, and deleting empty cells in their last 15 minutes. For participants who performed little notebook cleaning work at the end of the study, we asked why they chose not to do the cleaning work. We also asked about any cleaning work they would do if they were to share the notebook with others.

After the 15th interview, interested in what users valued in notebook work, we added questions about what they would teach to a novice notebook user.

\subsection{Task}\label{subsec:task}
%


Each participant worked on their own ongoing tasks during the study.
\ins{Prior to each study, we asked participants to bring their own tasks that could be completed within the time-frame of the study. Although the tasks required domain knowledge, each participant chose tasks in their own field of expertise.
As a result, all of the challenges we observed were computational rather than domain-specific.}
Most participants worked on a single task during the observational session, but three subjects (P8, P15, P20) completed their first task and started a second one. 
We coded each of the task descriptions 
to categorize them into seven different task types\ins{ based on our observations}: analysis, visualization, refactoring, reproduction, development, data cleaning, and notetaking. 
Reproduction tasks involved taking existing data and an expected output and recreating that output from the data in a new notebook.
We saw 23 tasks across 20 participants. All except P20 did the tasks in one notebook. 
\ins{To derive the approximate size of work in each notebook, we counted the scrolling activities that occurred during the study to approximate task size (defined as ``Scroll Size'' in }\autoref{tab:participants}\ins{) using the low-level action codes, the derivation of which we describe in }\autoref{subsec:data-analysis}\ins{.}
\autoref{tab:participants} shows the complete task descriptions\del{ along with }\ins{, }their respective task types\ins{, and the scroll sizes of their corresponding notebooks}. 


\subsection{Data Analysis}\label{subsec:data-analysis}




We open-coded the study recordings, which resulted in two subsets of \emph{low-level codes}: (1) those pertaining to actions conducted by the participant, and comments and preferences they indicated during the observational portion, and (2) those relating to the interview responses. 
\del{Two}\ins{The first two} authors\ins{, who also ran the studies,} coded the observational components of each study, and \ins{the third and fourth}\del{two other} authors coded the interviews\ins{ after reviewing the coded observations and field notes}. 
While coding the observational components, the \ins{first }two authors also coded \emph{meta-observations}, \ie, high-level observations they made about each participant's behavior implied by their actions or quotes.
The \del{authors}\ins{four author-coders met weekly to review the low-level codes until achieving agreement.}\del{ discussed and agreed upon the low-level codes.}

Finally, to further seek patterns among the low-level codes, \ins{the first }two authors conducted a second round of coding upon codes of participant comments, preferences, and meta-observations from the observational portion, and codes from the interview portion.
This round of coding is top-down as the coders derive \emph{high-level codes} implied by the low-level codes under three categories: quality attributes of notebook content, native notebook attributes related to the quality attributes, and user tactics. 
We then created memos to relate the user tactics to the quality attributes and the native notebook attributes, following the grounded theory practice. We categorized the tactics as either a direct use of notebook attributes or a workaround for notebook limitations.
We report the identified quality attributes, and their associated notebook attributes and user tactics in \autoref{subsec:quality-attributes}. We also include the full list of codes and a replication package in the paper supplement~\cite{Huang_Ravi_He_Tian_Lerner_Coblenz_2025}.


\subsection{Limitations}
While our starting question, \textit{how do people use Jupyter notebooks?} is broad, our participants primarily work in academia either as graduate students or as scientists. 
As a result, 16 of the 23 tasks were part of ongoing research projects\ins{, limiting generalizability to }\del{. Our results may have limited generalizability to non-academic and }non-research settings. Also, the projects we saw were of limited size; different techniques could be used by those who work on very large codebases.
\ins{Participants brought in their own tasks, which might result in varying task difficulties albeit improving external validity.}
\ins{A confirmatory survey and member checking could help further validate the findings, but the methods we used reveal the lived experiences of our participants}~\cite{yardley2008demonstrating}\ins{.}
We met with each participant once, so we were not able to see the notebooks they worked on evolve over time. 
Furthermore, our participants were all part of the same institution as the authors, limiting external validity. \ins{Representativeness may also be limited given that 15 of the 20 participants are graduate students, although many scientists who write code are graduate students}~\cite{hannay2009scientists}\ins{.}
\ins{Finally, five}\del{ out of 20}\ins{ studies were done in person instead of on Zoom, and the modality differences could have affected participants' performance.
}

Although the authors are computer scientists, researchers with other perspectives---particularly with more domain knowledge---could have had different understandings of the work we observed, and our software engineering perspective could have biased our interpretations of the efficacy of the tactics used by our participants. \ins{In addition, our study could be biased towards the fact that the participants prioritized Jupyter notebooks over other tools (\eg, MATLAB, command line tools) for computing in the observed contexts.}
%



\section{Results}
\label{sec:results}



We identified three categories of goals that participants had in their work, ranging from \emph{disposable exploration} to \emph{artifact construction} (\ins{\textbf{RQ1}, }\autoref{subsec:goals}). Participants valued eight \del{different }non-functional requirements, which we describe using quality attributes (\textbf{RQ\del{1}\ins{2}}, \autoref{subsec:quality-attributes}). They promoted those quality attributes using 18 tactics (\textbf{RQ\del{2}\ins{3}}, summarized in \autoref{tab:QA_summary}). In addition to our research questions, we report how participants used AI-based tools in their notebook work (\autoref{subsec:AI}).

\subsection{Participants}
We recruited 20 participants (11 identifying as male, 9 as female). Seven had prior work experience in software engineering, \chg{1}{one} had formal training in software engineering, \chg{6}{six} self-reported some familiarity with software engineering, and \chg{6}{six} reported none.
\ins{Half of the participants (11) were from computing-relevant domains (10 in Computer Science, one in Data Science) and nine from non-computing-relevant domains across Biology (three), Social Sciences (three), and Earth Sciences (four), where P6 had experience in both Social Sciences and Earth Sciences. 
}%
\autoref{tab:participants} displays the participants' backgrounds along with the tasks they worked on.

\begin{table*}[th]
\centering
\caption{Participant Backgrounds and Tasks. \\
\ins{In ``Field'', ``\comp''=Computing, ``\bio''=Biology, ``\soc''=Social Sciences, and ``\earth''=Earth Sciences.}\\
\del{Each participant worked on one notebook except P20. }``\&'' in Task Type denotes more than one task done in the study. \\ 
\ins{Each participant worked on one notebook except P20, as ``;'' denotes data for separate notebooks in Goals and Scroll Size. }\\
In Goals, ``A" represents \emph{artifact}, ``F" represents \emph{findings}, and ``DE" represents \emph{disposable exploration}.}
\label{tab:participants}
\begin{tabularx}{\textwidth}{llp{1.6cm}p{2.3cm}lXp{2cm}ll}
\toprule
\multicolumn{1}{l}{ID} & \multicolumn{1}{l}{Gender} & Occupation & Field & SE Experience & Task Description & Task Type & Goals & \ins{Scroll Size}\\ \midrule
P1                     & M                          & PhD Student                 & \ins{[\comp{}]}CS            & Work          & Refactoring a data analysis notebook                        & Refactoring  & A & \ins{33} \\
P2                     & F                          & PhD Student                 & \ins{[\comp{}]}CS            & Work          & Algorithm \& data comparison                               & Analysis & F                                                  & \ins{61}  \\
P3                     & M                          & PhD Student                 & \ins{[\comp{}]}CS            & Training      & Data visualization                                          & Visualization & A                                            & \ins{65}   \\
P4                     & M                          & PhD Student                 & \ins{[\comp{}]}CS            & Knowledge     & Exploratory data analysis                                   & Analysis     & F                                       & \ins{11}        \\
P5                     & M                          & PhD Student                 & \ins{[\bio{}]}Bioinformatics              & Knowledge     & Exploratory algorithm analysis                              & Analysis & F                                             & \ins{50}       \\ \vspace{.1em}
P6                     & F                          & Data Analyst \& Researcher & \ins{[\soc{}]}Economics \& \ins{[Earth]}Oceanography  &     Knowledge          & Refactoring a data analysis notebook                        & Refactoring                                                & A  & \ins{16} \\
P7                     & F                          & PhD Student                 & \ins{[\soc{}]}Neuroscience                & Knowledge     & Reproducing an existing notebook                            & Reproduction              & DE                  & \ins{17}   \\
P8                     & M                          & MS Student                  & \ins{[\comp{}]}CS            & Work          & Data cleaning \& developing a machine learning model       & Data Cleaning \& Development & F                            & \ins{47}   \\
P9                     & M                          & Scientist                   & \ins{[Earth]}Geoscience                  & None          & Reproducing data visualizations                             & Reproduction & F                             & \ins{12}     \\
P10                    & M                          & Scientist                   & \ins{[\bio{}]}Microbiology                & None          & Migrating a script to a notebook for documentation & Refactoring & A                                      & \ins{71}        \\ \vspace{.1em}
P11                    & F                          & PhD Student                 & \ins{[Earth]}Oceanography & None          & Data visualization homework assignment                      & Visualization  & A                   & \ins{35}        \\
P12                    & M                          & PhD Student                 & \ins{[\comp{}]}CS            & Work          & Code cleanup                                                & Refactoring  & F                                       & \ins{24}        \\
P13                    & F                          & Lab Assistant               & \ins{[\soc{}]}Psychology                  & None          & Data visualization                                          & Visualization & DE                                    & \ins{13}          \\
P14                    & F                          & PhD Student                 & \ins{[\comp{}]}CS            & None          & Algorithm implementation                                    & Development & F                                      & \ins{9}          \\
P15                    & F                          & PhD Student                 & \ins{[Earth]}Geoscience                  & Knowledge     & Data analysis \& visualization                             & Analysis \& \hspace{1em} Visualization   & F   & \ins{21}   \\
P16                    & M                          & PhD Student                 & \ins{[\comp{}]}CS            & Work          & Analysis of machine learning models                         & Analysis   & A                                        & \ins{32}         \\
P17                    & M                          & PhD Student                 & \ins{[\comp{}]}CS            & Work          & Data analysis                                               & Analysis   & A                                       & \ins{13}          \\
P18                    & F                          & PhD Student                 & \ins{[\comp{}]}CS            & Work          & Testing different machine learning models                   & Analysis  & F                                  & \ins{35}                \\
P19                    & M                          & Undergraduate       & \ins{[\comp{}]DS}\del{Data Science}                & Knowledge     & Drafting a programming \hspace{1cm} assignment                           & Development         & DE                           & \ins{30}             \\
P20                    & F                          & PhD Student                 & \ins{[\bio{}]}Bioinformatics              & None          & Annotating a notebook \& \hspace{0.5cm} reproducing visualizations (two notebooks)         & Notetaking \& Reproduction  & DE\del{,}\ins{;} A            & \ins{3; 5}      \\ \bottomrule
\end{tabularx}
\end{table*}


\subsection{Goals for Using Jupyter Notebooks}\label{subsec:goals}


11 codes pertained to the kinds of goals that participants had in their work. For example, some participants described focusing on scientific findings, whereas others focused on presenting their work to others. We also identified 15 codes relevant to ensuring the clarity of a notebook. 
These codes include five categories: removing redundant code and cells, taking notes (as comments or Markdown), refactoring code (\eg, renaming variables), inserting empty cells to separate sections, and reformatting or editing code for code and/or output readability.
We also noticed varying expectations for notebook longevity---\ie, whether to revisit the notebook after the task demonstrated in the study is done---within the first ten participants, and we started asking about the expectations for notebook longevity explicitly since P11. 

Combining these observations, we found that each task fit into one of three categories: \emph{disposable exploration}, \emph{findings}, and \emph{artifact}.
%
%
\emph{Disposable exploration} refers to exploratory work that will be discarded immediately after the outcome is achieved. 
We consider a notebook used for disposable exploration if it does not have any expected longevity.
An \emph{artifact} details the process and outcome of problem solving and/or scientific discovery in a clear, descriptive, and potentially reproducible way.
We consider a notebook to be an \emph{artifact} if the task where it is used is a cleanup task, or it has expected longevity and its author showed three or more kinds of the clarity-related actions (\ie, more than half of the five available kinds, to demonstrate sufficient effort in ensuring clarity from multiple aspects). 
%
%
Finally, a \emph{findings} notebook documents the process and outcome of problem solving and/or scientific discovery but not necessarily in a structured way---its main purpose is to expose information to the notebook author for them to decide on next steps of work.
Although a \emph{findings} notebook has some expected longevity,  its author used less than half of the possible kinds of clarity-related actions.

Out of 20 participants, we have 21 notebooks; P20 worked on two notebooks during the study, which we denote as P20A and P20B as necessary.
We found four notebooks for disposable exploration, nine for findings, and eight for artifact.
T\ins{he first t}wo authors compared the results with each individual study and field notes and agreed upon the categorization.
\ins{While only P20 worked on more than one notebook during the observations, several participants (P3, P5, P11) shared the context of their notebook work via multiple notebooks prior to the observation or during the post-observation interview.}


\subsection{Software Quality Attributes}\label{subsec:quality-attributes}


Participants valued eight quality attributes: clarity, reproducibility, explorability, debuggability, reusability, correctness, performance, and collaboration. We detail each quality attribute, how notebooks promoted inhibited quality, and user tactics that promoted quality (results summarized in \autoref{tab:QA_summary}).


\begin{table*}[tb]
    \caption{Summary of how notebooks affect quality attributes and tactics participants used to improve quality}
    \label{tab:QA_summary}
    \centering
    \begin{tabularx}{\textwidth}{lp{5.5cm}p{4.5cm}X}
        \toprule
          \textbf{Attribute} & \textbf{Ways notebooks inhibit} & \textbf{Ways notebooks promote} & \textbf{Tactics used to promote} \\
          \midrule
         Clarity & Flat, cell-based structure of notebooks makes it difficult to organize information & Markdown notetaking & Abstraction; sectioning; maintaining one task per cell; creating new notebooks \\
         Reusability & Single scope results in accidental variable reuse & Single scope avoids need for parameter passing & AI-based explanations of unfamiliar code\\
         Reproducibility & Out of order cell execution; no built-in package management & Broad usage of Jupyter enables viewing and running others' notebooks & Virtual environments; re-running from the top; readability and cleanliness; storing notebooks alongside data\\
         Explorability & Lack of support for output comparison across runs; information overload; manual cell execution and state management; inability to inspect in the middle of a cell or in the middle of a loop & Cell structure-based interactions and code-ouptut correspondences & Writing intermediate outputs to disk; exploring new ideas in fresh, short notebooks; merging cells with state dependencies\\
         Correctness & Error-prone manual state management & Cell-by-cell execution allows users to check the validity of each line & Cell-based risk management; restart and run all; enforcing linear execution \\
         Performance & Error-prone manual state management & Caching data; saving outputs; cell-by-cell execution & Reusing data; sectioning \\
         Debuggability & Out of order cell execution; single scope when debugging inside a function; difficulty with navigating to relevant buggy cells; inability to inspect in the middle of a cell; enforced kernel restart with changes in dependencies & Cell-based structure promotes small inspection, code-output correspondence, and rapid edit-run cycles & Restart and run all; avoid debugging within function definitions; notetaking \\
         Collaboration & Limited compatibility with file diffing utilities & Broad usage of one notebook tool (Jupyter) makes collaboration easier & Sectioning \\
         \bottomrule
    \end{tabularx}

\end{table*}

\mypara{Clarity}
17 participants described tactics that they used to improve the clarity and presentability of their notebooks. 
Ten of our participants 
planned to present their work to their colleagues using their notebooks, so they took extra care to ensure that the notebook was readable and could even be edited and recomputed on the fly. 
Others, including P17, P18, P19, and P20 reported that they often refer back to code written in previous notebooks and needed to be able to understand and potentially reuse code from them. 
P17 described reusing code from notebooks dating back to 2018 and said he would continue writing new code in the notebook in the future.

Seven of our participants highlighted that they appreciated the ability to interleave Markdown notes with their code. 
P10 said that these notetaking abilities in notebooks make it easier for others to understand the code in notebooks, and P20 stressed that \textit{``the markdown function is extremely important to [her]"} when writing exploratory code. 
For some users like P1, too many notes can have the opposite effect and hinder clarity. Instead, he wrote high-level to-dos elsewhere. 
P7 had a similar strategy and explained that she writes notes in a separate notetaking software because it was better at tracking history compared to notes inside the notebook.
The Markdown capabilities of notebooks also allow users to create sections in their notebooks by creating headings for certain groups of cells, which both P1, P2, and P20 used. 
Alternatively, P9, P16, and P19 split up their notebooks using multiple empty cells to segregate tested code from exploratory code and to separate different paths of exploration. 

Over half of the participants used the flexibility of the cell structure to organize content for clarity.
They adopted the heuristic of ``one task per cell" to keep relevant lines of code together while maintaining the ability to see the outputs of intermediate computations; however, P1 noted that there was a tension between wanting to group related code and wanting to break apart and inspect inside a cell: \textit{``On one hand, I want the flexibility to be able to look inside a cell and really get into its pieces. But, on the other hand, I also want to be able to flip it over and be like, okay, I've iterated on some kind of structure, and I have this modular building block."} 

This form of content organization aids the reuse of code across notebooks since notebook code cannot be exported and must be reused via copy/paste. 
However, copy-pasting code can impede readability when unnecessary or redundant code is added to a notebook. 
P18 found some vestigial code copied from another notebook and noted \textit{``sometimes I'll copy and paste old code into here and then I'll just forget to delete it."} 

The flat, cell-based structure increased the participants' cognitive load as inspection code and outputs interleaved.
P16 remarked that \textit{``once there are too many things [...] happening in a notebook it becomes hard to follow."} 
To manage information load, P14 would reuse cells for multiple inspections, and both P13 and P16 would delete inspection code to reduce visual clutter. P16 also preserved inspection code in comments for later use to avoid retaining the inspection output. 


Five of our participants (P12, P16, P17, P18, P20) reported that they  
create new notebooks to explore new ideas, debug, and clean up code, which required copy and pasting code from notebooks. 
P17 explained that he sometimes creates a new notebook and copies over his code cell-by-cell to debug; P2 completed the study using a notebook she created exclusively for debugging. 
P5 inspected a dataset in an existing, cleaned-up notebook instead of in his current notebook. 

Two participants wrote code in functions to explicitly promote clarity in their notebook. 
P18 explained that she abstracts code into functions to help her focus on relevant information while reading through her notebook.
P14 explained in her interview that \textit{``at the end it's nice to have a bunch of functions when the code is cleaned,"} but like P1, she preferred to lose the clarity of functions when developing to ease debugging. 


\begin{tcolorbox}
\textbf{Takeaway 1}: Users put effort into ensuring notebook clarity and often use Markdown cells in Jupyter. Whereas typical software engineers rely on abstraction and structure to make code understandable, notebooks' flat structure does not provide these capabilities and often hinders clarity.
\end{tcolorbox}

\mypara{Reusability} 
Traditionally, software engineers leverage modularity and abstraction to promote reusability~\cite{SAIP}, since abstractions can be reused across contexts without understanding modules' implementation details. In contrast, Jupyter's flat namespace and single scope for all cells both facilitated reuse (by avoiding the need to pass parameters or change representations) and inhibited reuse (by enabling bugs caused by variables having meanings that pertain to irrelevant parts of the program). Participants reported writing functions when code would otherwise be duplicated, but P3 and P11 preferred to duplicate code unless it would result in more than several copies; P11 explained that repetitive code can be easier to read than non-repetitive code that invokes functions. 


Jupyter's single scope caused problems for P12, who copied and pasted code within a notebook but forgot to update a variable, which was still bound due to its previous use. Fortunately, after seeing plot emitted by a cell, P12 debugged and fixed the problem. Later, P12 encountered another instance of the same problem, but did not notice the bug.

Because of mutation, code reuse within a notebook even for the same purpose is unsafe. Participants often invoked Pandas functions that, for example, renamed columns, so code that is correct before the renaming operation would be incorrect afterward. P1 became unsure which lines of code would change the structure of a dataframe, restarting evaluation to be sure: \emph{``okay, let's start from the top.''} 

Some participants wanted to reuse code between notebooks and Python scripts. For example, P10 adapted code from a standalone script for use in Jupyter. However, Jupyter couldn't invoke the script's \texttt{main} function. P10 refactored the code to use variables instead of command-line arguments. P7 worked with example code that reflected this same pattern: it included a function called \texttt{mainfunction}.

Some participants wanted to reuse \emph{unfamiliar} code from other contexts, which required understanding the code to be reused---at least to some extent. P7 relied on ChatGPT to explain unfamiliar code from an example that she wanted to reuse, even though the code included various comments. But these tools were not integrated into Jupyter, so P7 had to copy and paste the code into another window, leaving ChatGPT without the code's surrounding context. P7's query to ChatGPT was only the source code followed by \texttt{explain}, missing a possible opportunity to ask a more specific question.

P16 described converting notebook code to scripts, which execute outside Jupyter; this process is facilitated by the fact that Jupyter provides only a thin interface on top of Python. 

\begin{tcolorbox}
\textbf{Takeaway 2}: Jupyter's lack of abstraction promotes frequent copy/paste. The flat namespace and single global scope makes copy/paste convenient but error-prone.
\end{tcolorbox}

\mypara{Reproducibility}
Reproducibility concerns the ability for the developer or others to reproduce the same notebook output in the future. Reproducibility is important in replicating scientific analyses and extending prior work with new analytic techniques. Unfortunately, reproducibility can be a real problem for notebook users. P17 recounted a situation in dealing with a non-reproducible notebook: \emph{``I simply created a blank notebook and copied section by section, because I think \emph{this} section would run […] if it does run, then I move on to the next section, and that does identify the problem.''}

Because notebooks permit out-of-order cell execution, re-executing notebooks in order can produce different results than users first observe. At the end of each session, we asked participants to re-run their notebooks in order. P2 and P9 were unable to reproduce their earlier work this way, suggesting that out-of-order execution is a real threat to validity. Some participants (e.g., P3) were careful to write code in order of dependencies, but this process was manual. P6 and P8 used ``restart and run all'' to make sure their notebooks would run in order; P17 complained about how out of order execution threatens reproducibility. P16 cited mutation and order of execution when explaining why he kept imports at the top.

P1, who had a computing background, used a package manager, Poetry, to create virtual environments for notebooks, enabling specification of dependencies. In contrast, most participants did not appear to be concerned with library versioning, which could threaten reproducibility. 

Nine participants considered reproducibility to include readability and cleanliness, since readers might need to understand the code to reproduce the analyses. Matters of readability are discussed under the \emph{Clarity} heading in this section.

\begin{tcolorbox}
\textbf{Takeaway 3}: Scientists value reproducibility, but out-of-order execution and lack of package management hinder~it. 
\end{tcolorbox}


\mypara{Explorability}
Exploratory programming is about prototyping ideas and iterating on implementations through code without pre-defined specifications or goals~\cite{Kery2017:Exploring}.
Regardless of their goals, our participants valued explorability; indeed, explorations are prevalent in programming, and even the process of creating an \emph{artifact} involves exploration. Jupyter includes features that promote exploration: cell-based interactions, facilitating exploration through small inspection (P8, P11, P17), nonlinear execution (P20B), and interleaving code and output for correspondence and quick comparison (P12, P17).

Participants often needed to compare outputs between different versions of their code, but Jupyter did not facilitate this: every time they re-ran a cell, the old output was overwritten. To work around this problem,
%
P1 saved output outside of the notebook for comparison: \emph{``I [would] have code blocks output their result to a file or […] save it somewhere […] and then I'll copy paste that result into like a constant in the code block [for comparison].}
P8 chose another approach, leveraging out-of-order cell execution: putting various implementations of an algorithm in different cells enabled comparing the outputs.

Notebooks truncate cell output if it is too long, even though the output could include important information that is easily missed.
P9, for example, missed a message indicating an installation failure because it was buried in a long output, instead believing that installation had succeeded. This caused persistent failures when he ran other cells that used functions in that package.
Long outputs can also cause difficulties when testing out multiple new ideas by reducing clarity. 
As a workaround, P12 and P16 started fresh notebooks for exploring new ideas to avoid a notebook getting long due to too many inspection cells and outputs, \emph{``just [making] a new notebook [...] if something gets messy''} (P12). 

Jupyter requires users to manually rerun a cell after it has been edited, but users sometimes forgot to do so, leading to unexpected output and changes to the global state.
These changes made it difficult to assess the validity of exploratory code. 
For example, P2 changed the input file she was using to a truncated version in order to do more explorations. 
However, she forgot to rerun this edited cell and operated under the impression that she was working with the truncated files. 
This caused her to both waste extra time waiting for the runs to finish and created an extra bug for her to solve, distracting her from her original task.  
%
P8 tried to avoid such hiccups by packaging exploratory code into a function with exploratory parameters as the arguments and putting a call to that function in the same cell, so that he could repeatedly call the function with different parameter values to explore outputs.
%

Finally, similar to how they used markdown notes and annotations to keep notebook content clear, participants used notetaking to facilitate explorations so that they could freely explore without getting lost. 
This way, P20 said, \emph{``I [could] know that [which] is the the newest exploration [...] and these are all the file paths [...] that I wanted to use [for it].''}

Jupyter's limitation of only showing values of expressions that are at the \emph{ends} of cells frustrated P13, who expected to inspect an expression in the middle of a cell without using the \texttt{print} function. Likewise, the promise of expression-based interactivity enabled by the cell structure breaks with loops: P4 could not inspect expressions within a loop unless printing them out and was forced to rerun the whole loop, as opposed to individual iterations, to gain any feedback on code change.

\begin{tcolorbox}
\textbf{Takeaway 4}: Cell-by-cell execution can help users iterate in straight-line code, but the cell model has difficulty scaling to more complex workflows. 
\end{tcolorbox}

\mypara{Correctness}
In traditional software engineering, systematic testing is used to evaluate the correctness of code. In contrast, our participants found it difficult to concisely describe expected results. 
P17 notes that \textit{``it's not like you can write unit tests to see if things are correct, sometimes you can tell by the data. If the data distribution doesn't look right, then [I] realize maybe I should have done it differently."} 
Lack of functional decomposition makes it difficult to write unit tests: only five out of 20 participants wrote new functions.
In addition to inspecting output manually, participants adopted notebook-related strategies to leverage the notebook environment and mitigate its risks, including \emph{cell-based risk management} (described below), \emph{enforcing linear execution}, and \emph{restarting and rerunning notebooks} for ensuring the correctness of their code. 

In \textit{cell-based risk management}, used by four participants (P1, P2, P8, P17), users manage the risks presented by new code by writing in separate cells, which they later combine.
For example, P8 created a new cell to remap a categorical column of a dataframe to numerical values. 
He checked that his code worked as expected by inspecting the datatype of the column. 
As he needed to do the same for three other columns, he created a new cell and wrote similar code for all three remappings in the same cell and ran the cell without additional inspection. 
When asked why he chose not to inspect, he said 
\emph{``I did my proof of concept for the first thing... I know it's going to work because it worked for one of them.''}

Because Jupyter requires users to manually manage state, some participants took time to ensure that they had the correct mental model of their notebook's execution.
When trying to understand a collaborator's notebook, which involved a lot of variable mutation, P1 said \textit{``I get nervous about this stuff because I don't know if I've reset the state. So, my way of handling that is just restart and run it from the top."}
P6, a newcomer to notebooks and programming, runs each of her cells again whenever she makes a major change in her code to ensure that there are no new errors. 
Others (P2, P3, P8, P12, P17) enforced linear execution in their notebooks.

Restarting and running all notebook cells is not always desirable, especially when working with large data. P4 said, \textit{``the nice thing about Jupyter [...] is like just loading the data and not having to load it every single time when I run a script."}
Restarting and running all the cells to ensure correctness in this case would counteract the performance benefits of notebooks. 

\begin{tcolorbox}
\textbf{Takeaway 5}: Traditional software testing methods are difficult to incorporate in notebooks, so users mitigate risk through inspection and cell-by-cell execution. 
\end{tcolorbox}

\mypara{Performance}
Some participants praised notebooks for facilitating working with large datasets. P4 talked about how his dataset takes 2-3 minutes to load, but using Jupyter allows him to just load it once and run his computations as many times as he needs. 
P2 explained, \emph{``the point of the Jupyter Notebook is that I have [computations] saved so I can use them later.''}

Participants promoted performance in their notebook work by limiting the number of times they loaded data and ran unnecessary computations.
Out-of-order execution and splitting up cells allow both reusing loaded data and running the code efficiently. 
P11 preferred to complete multiple tasks in the same notebook and created sections to separate them. To run the code for a single task, she would run the first notebook cell containing all the import statements she needed, and skip to only the cells in the relevant section.

Initially, P2 wrote code following the ``one task per cell" principle. However when debugging code that took a long time to run, she split up her cells based on how often she needed to recompute certain lines and how long they took to run. 
Cycles of editing and rerunning the split up cells caused some confusion about the current state of the notebook and whether certain cells had been ran after changes. Ultimately, P2 had to rerun each of the expensive computations again in order to confirm that she was working with the most up-to-date outputs. 
This tactic that was intended to aid performance ended up hindering it when used in the context of debugging. 



\begin{tcolorbox}
\textbf{Takeaway 6}: Notebooks benefit data-heavy tasks by enabling partial execution of programs, but users must carefully manage state to leverage this feature.
\end{tcolorbox}

\mypara{Debuggability}
Debuggability refers to the ability to determine the cause of a bug.
All tasks but cleanup tasks P6 and P20A involved some debugging. The cell-based model enabled users to see output of small portions of the program, making the edit-run cycle much faster than in traditional IDEs. Cells enabled participants to compare and connect code to output (P12), run cells out of order (P2, P8, P13), and isolate code for debugging errors (P9, P19). However, out-of-order execution, single global scope, difficulty in finding relevant cells, requiring reloading when dependencies are changed, and expression-based inspection having to be at the end of a cell all interfered with participants' ability to debug efficiently. 

Out-of-order execution in notebooks required the user to manually rerun cells that had been edited (and all other cells that depended on them). 
For example, P15 had written a \texttt{for} loop that was supposed to update values in rows of a Pandas data frame. Unfortunately, P15 neglected to index \emph{into} the frame, accidentally rewriting the entire column in every iteration.
The first iteration ran with some output, but the second iteration failed due to the unexpected change in the entire column.
Confused, P15 decided to rerun the cell to replicate the error, only to see the \texttt{for} loop fail immediately during the first iteration: it was now operating on data that had been mutated in the previous (failed) run. 15 minutes into the situation, the participant sought help from the interviewer, who explained that P15 had to reload the data frame, resolve the bug in the loop, then rerun the cell with the loop to finally see the expected output.
Neglecting an index may be common in dynamically-typed languages, but a scripting setting would not have produced the output that misled the participant across runs as every run executes the entire script, not just snippets of code.
Like P1 pointed out, one must resolve some debugging scenarios in Jupyter notebooks by restarting the kernel and rerunning all cells to enforce the script-like execution linearity.

The single global scope in notebooks also makes debugging and inspecting local variables in a function hard (P3, P10).
For example, P10 considered returning local variables he wanted to inspect from a function 
to use the expression-based inspection in a cell.
Instead, P14 simply avoids debugging inside a function. 
When processing large files inside a function, the edit-run-inspect cycles can take a long time. By moving code out of a function, she could see intermediate output without having to stop, add \texttt{print} statements, and rerun the code.


Compared to an IDE for scripts, where one could easily go to the definition of a function or simply a specific line of code to localize the bug, notebooks provided no easy way to navigate to relevant cells (P3, P9, P12).
In particular, P9 could not locate the cell he just ran after he scrolled through the notebook to read other cells while waiting for the execution to complete.
As such, participants spent a lot of time scrolling through the notebook: in fact, scrolling was the most prevalent action across all studies, with 603 coded instances out of 4195 total action instances (14.3\%).
To complement the lack of navigation aid in notebooks, participants (P14, P16) took more notes in the notebook to facilitate cell navigation in debugging: \emph{``If you leave notes [...] then if something doesn't work, at least I can go back and look at my notes, [...] start with the things that looked weird intermediately, and go from there''} (P14).

Notebooks required restarting the kernel for changes in the dependencies to take effect, which severely slowed down the edit-run debugging cycles for P7 when some of the debugging-related code changes occurred in an imported module.

Jupyter shows the value of the last expression in each cell. P13 wanted to inspect arbitrary expressions in cells without inserting \texttt{print} statements. Enabling easy inspection of all values could further promote debuggability.

\begin{tcolorbox}
\textbf{Takeaway 7}: Cell structure helps isolate errors, but out-of-order execution and single scope impedes debugging because debugging work can mutate state needed elsewhere. 
\end{tcolorbox}

\mypara{Collaboration}
Collaboration is an important part of the scientific process. As students and researchers, our participants needed to ensure that their work could be understood by others, and when needed, could be collaboratively written.
Two of the twenty participants planned to co-author their notebooks with their colleagues (P6, P14), but many planned to share and iterate on their notebook work with the input of others (P1, P3, P5, P10, P11, P12, P16, P17). 
However, Jupyter does not natively provide many tools for facilitating collaboration on the same notebook, so users rely on ad hoc methods such as splitting a notebook into separate sections and storing the notebook on a shared drive.  
Version control systems have limited benefit because they do not integrate nicely with notebook cells. This also affected our participants' willingness to collaborate with others on notebooks. P8 said that he collaborates with others when working with Python scripts, but chooses not to collaborate with others because of the difficulty of resolving editing conflicts in notebooks. 
To avoid messy conflicts, P20 elaborated in interview that she had once split a notebook into two sections by adding empty cells in the middle, and she and her collaborator worked on the cells on opposite sides of the divide. 

\begin{tcolorbox}
\textbf{Takeaway 8}: Scientists often work together, but a lack of version control integration or other collaboration tools for notebooks makes collaboration difficult. 
\end{tcolorbox}

\subsection{Use of AI tools}
\label{subsec:AI}
11 participants used AI-generated code during their respective studies. Three (P1, P2, P5) used GitHub Copilot \cite{friedman2021copilot}, which was integrated into their notebook environment (\ie, VS Code). Nine participants (P5, P7, P9, P10, P13, P15, P17, P19, P20) used ChatGPT for help when writing notebook code, and two (P12, P16) mentioned using it in the past.

The participants using Copilot largely used it as an autocomplete tool to accelerate their productivity \cite{barke2023grounded} since it was \emph{``a good time saver,"} (P2).
In 18 of the 21 instances of Copilot usage, Copilot would complete their line of code, and 12 instances were accepted without changes. Two were considered unhelpful after reading and were deleted, and three were accepted and edited. 
P1 also used Copilot to explore solutions by prompting it in the comments to create a plot in a new cell. Once he ran the cell, he saw that the axes labels were unreadable. He tried to fix this issue by prompting Copilot to fix the code, but when unsuccessful, moved on. 

\autoref{tab:ai} shows 14 instances ChatGPT usage, 11 of which had a participant successfully integrate ChatGPT-generated code into their notebook.
To validate the AI-generated code, each of these participants first read over the code, and all but one (P5) copied and ran the code inside their notebook to determine if it met their needs. 
However, this strategy did not always succeed: P9 could not download the dependencies needed to run the generated code. 
In addition, P7 relied on ChatGPT twice to explain code from the notebook, but noted \emph{``I don't know this code well enough to tell if ChatGPT is giving me something wrong.''} 
All but one ChatGPT user worked in a non-computer science field, which highlights the important role of AI tools for scientists in programming.

The exploratory nature of notebooks eases validating AI-generated code because, as P12 puts it, \emph{``if it gives me code that might be wrong, I can just try it."}
Still, for those with less coding expertise, it may take them more effort to integrate the code into notebook before validation by execution is possible.

\begin{tcolorbox}
\textbf{Takeaway 9}: Scientists rely on AI-based tools even though they are not integrated into their environments, but sometimes lack the programming knowledge to understand or incorporate AI-based suggestions successfully. 
\end{tcolorbox}


\begin{table}[tb]
\small
\centering
\caption{Usage of ChatGPT}
\label{tab:ai}
\begin{tabular}{@{}llll@{}}
\toprule
ChatGPT Usage Scenario & Participants & Success Rate \\ \midrule
Explaining code   & P7  & 2/2  \\
Fixing errors  & P13, P17 & 2/2  \\
Generating visualization code & P9, P13 & 3/3 \\
Using unfamiliar libraries & P5, P9, P15, P19 & 1/5 \\
Other programming tasks & P10, P13, P20 & 3/4 \\
\bottomrule
\end{tabular}
\end{table}
\section{Discussion\ins{ and Future Work}}

\ins{
Identifying quality attributes that \emph{scientists} value exposes opportunities to build theory and deepen our understanding of prior work, which focused on \emph{data workers} more generally.
First, our study provides a framework for understanding notebook use and challenges in scientific settings through the lens of quality attributes and identifies tactics scientists use to promote the reported quality attributes.
This could enable theory-building opportunities; data collection could focus on the root causes of priorities and motivations, which would enable a stronger theoretical perspective. 
For example, our study identified conflicting quality attributes for certain goals and contexts (\eg, clarity could conflict with debuggability when a notebook aims to show more intermediate computations), and in-depth data collection could better explain how scientists navigated quality attributes in conflict.
Second, with a specific focus on scientific users, while our study reveals notebook usage goals and pain points similar to prior results (of data workers)}~\cite{kery2017variolite, subramanianCasualNotebooksRigid2020, Chattopadhyay2020:Whats, ruleAidingCollaborativeReuse2018}\ins{, we found that priorities for notebook quality attributes depended on context and goals (rather than a general prioritization of exploration over explanation).
}

\ins{We noted highlights in the current design of notebooks that supported scientific work and its associated quality attributes. }Support for dividing code into cells that can be executed in any order and share an execution environment is the hallmark of the computational notebook paradigm. As we observed, this promotes explorability, which our participants prized. Combined with the support for Markdown-based headings and explanations, which promotes clarity, computational notebooks focus on exactly the quality attributes most valued by our participants. Support for easily seeing output also promotes correctness and debuggability, and out-of-order execution facilitates work with very expensive analyses (performance). Our participants were less concerned with reproducibility and reusability, which are weaknesses of computational notebooks. 

How, then, could future tools for scientists do better? Key design opportunities pertain to reusability and reproducibility, which are inhibited by out-of-order execution, the global scope for all cells, and lack of built-in package management. Here, we believe an opportunity lies in enabling more separation between the cells. Cells could have their own scopes, with explicit control over which variables are imported from and exported to global scope. This could enable a kind of reactivity, similar to spreadsheets or Observable~\cite{Observable}, in which cells are automatically re-evaluated when their inputs change.

Similar to that shown in other work~\cite{Rule2018:Exploration}, we observed a spectrum from exploratory work to explanatory work. Systems need to support work that spans this spectrum over time, but these are sometimes in conflict: out-of-order execution, which promotes explorability, inhibits reproducibility; also, single scope promotes explorability but inhibits some aspects of reusability. Also, some tactics promote some quality attributes at the expense of others, leading to refactoring needs when goals change. For example, a very exploratory notebook might include a large number of short cells to enable quick edit/debug cycles; as the notebook becomes more explanation-oriented, we found that participants were more likely to combine related code into larger cells. Future notebook tools could provide refactoring tools that make transitioning between goals more convenient, when users are ready---and even enable \emph{backward} transitions. Existing tools enable splitting cells, which helps, but documentation can increase viscosity~\cite{blackwell2001cognitive}; tools that track a tighter relationship between code and documentation and between different regions of code could make it easier to revise mature code. In our study, we repeatedly observed participants starting over with fresh notebooks when they needed to make these kinds of transitions, creating a mess of different files with unclear histories and necessitating re-work to construct each new version.

\paragraph{Modular notebooks} Notebooks have a linear structure: a sequence of code cells, interspersed with Markdown cells. The result is that notebooks \emph{hide dependencies}, inhibiting modularity~\cite{blackwell2001cognitive}: code cells can rely on variables that were bound or whose values were mutated in other cells in the notebook. Then, reusing a cell requires first identifying its dependencies. Tools could help find required code~\cite{Head2019:Managing}, but it could be more effective to promote modularity. Dependencies between cells could be restricted; lexical and semantic dependencies could be made explicit (adding annotations of preconditions and postconditions on cells~\cite{Hoare1969:Axiomatic}). Gradual approaches could promote formal abstraction mechanisms: refactoring tools could make it easy to extract cells as functions, and notebooks could adapt tools like Projection Boxes to enable live exploration of function behavior~\cite{lerner2020projection} (which Engraft~\cite{Horowitz2023:Engraft} has attempted). Another axis of modularity concerns data: code that reads from files or Internet data sources can be buried in notebooks, again hiding dependencies. Notebook environments and languages could improve modularity by making these dependencies, including requirements regarding file formats, explicit.

When executing a notebook, users can choose to run all cells, a single cell, or all cells up to a point. But this inhibits explorability because executing all of these cells can be very expensive. Alternatively, running just one cell is risky because the environment does not track which cells need to be re-run after recent changes. Modularity could enable analyses that let users explore more safely and efficiently.

\paragraph{Organizing large notebooks}
Cells in notebooks are in a fixed, scrollable view, making it difficult to see portions of the notebook that pertain to specific tasks. Other tools that facilitate data analysis are more flexible. Spreadsheets include multiple sheets, enabling users to divide analyses into sections. LabVIEW~\cite{LabVIEW}, a graphical programming environment that targets scientists and engineers, breaks projects into separate files and libraries. Although Jupyter notebooks can call functions in other notebooks via the \texttt{\%run} command, doing so pollutes the namespace with all of the referenced functions. Future computational notebooks could be more flexible, allowing users to organize their cells according to their content.

\paragraph{Parallel evaluation} Some cells can take a long time to run. When those cells are executing, no other cells can be evaluated. This restriction relates to notebooks' inability to analyze dependencies: from the environment's perspective, any cell can produce state that is needed for any other cell. But in general, this is not the case, and restricting progress in an unrelated part of the notebook inhibits exploration. A few notebook improvements have adopted dataflow analysis to report or indicate dependencies across cells~\cite{macke2021fine, koop2017dataflow}. 
Based on these works, we could identify non-dependent cells to parallelize their evaluation to address this problem.

\paragraph{Caching partial results} Cells can include some very expensive lines of code, whereas other lines of code are very cheap. Better analysis of dependencies in notebooks~\cite{koop2017dataflow, macke2021fine} could enable caching partial results, promoting exploration.

\paragraph{Packaging dependencies} Notebooks often depend on installed packages and data files. Current notebooks do not support specifying dependencies and their versions in metadata. Although users can work around this via virtual environments, notebooks could make this easier with explicit support. Likewise, users must manually bundle input data with notebooks, something that automated support could help with.

\paragraph{Better AI Integration}
Our study shows how scientists used AI assistants in notebook work.
Our findings align with prior evidence of AI use in IDEs~\cite{barke2023grounded, ferdowsi2024leap} and further unveil the need for built-in AI support in notebook environments, which prior work has proposed for general notebook use~\cite{mcnutt2023}.
Specific to scientists, future AI integrations in notebooks should emphasize supporting the \emph{validation} of AI-generated code~\cite{ferdowsi2024leap} (\ie, checking if the generated code matches one's intent) while taking into account their programming and domain expertise.

\section{Related Work}

Since Knuth's proposal~\cite{knuth1984literate}, numerous literate programming environments have enabled end-users to incorporate more storytelling~\cite{granger2021jupyter} into their code~\cite{wilkin1998software, hayes1990thoughts, baumer2015r, Perez2007:Ipython, granger2021jupyter, Observable}.
Jupyter notebooks allow quickly prototyping ideas through  cells and interleaving narratives with code, helping document  scientific discoveries and analyses~\cite{granger2021jupyter}.
Indeed, among the long line of computing environments scientists use~\cite{hayes1990thoughts, matlab2012matlab, bezanson2017julia, baumer2015r, spyder, LabVIEW, Observable, Perez2007:Ipython, granger2021jupyter}, Jupyter notebooks~\cite{Perez2007:Ipython, granger2021jupyter} have become very popular~\cite{Pertseva2024:Theory}.
\mypara{Corpus Studies}
With millions of public Jupyter notebooks on GitHub~\cite{granger2021jupyter}, multiple corpus studies~\cite{Rule2018:Exploration, Pimentel2019:LargeScale, liuRefactoringComputationalNotebooks2023a, grotov2022largeScale, Raghunandan2023:Code, DeSantana2024:Bug} gained insights into the usage patterns in Jupyter notebooks via such data.
These studies identified the tension between exploration and explanation in constructing and sharing notebooks~\cite{Rule2018:Exploration}, their lack of reproducibility~\cite{Pimentel2019:LargeScale}, and the lack of good coding practices in notebooks~\cite{grotov2022largeScale}.
Building upon prior results, recent work proposed a linear regression model to predict the level of exploration vs. explanation in a notebook~\cite{Raghunandan2023:Code}, developed a taxonomy of bugs in notebooks~\cite{DeSantana2024:Bug}, and analyzed refactoring behavior across the evolution of notebooks~\cite{liuRefactoringComputationalNotebooks2023a}.

These studies show that while notebooks can evolve from exploration-focused to explanation-oriented by introducing more clarity~\cite{Raghunandan2023:Code}, and notebook users do attempt debugging~\cite{DeSantana2024:Bug} and refactoring~\cite{liuRefactoringComputationalNotebooks2023a}, notebook code may be of low quality according to traditional metrics, such as presence of unused module imports~\cite{Pimentel2019:LargeScale, grotov2022largeScale}.
These quality metrics, however, may be less relevant for scientists than the higher-level quality attributes that we identified.
%
In addition, these studies analyze notebooks on GitHub, which might miss some insights since scientists often choose not to publish notebooks on GitHub except for sharing~\cite{Pertseva2024:Theory}.
Our study revealed notebook-specific quality attributes that scientists valued, tactics they performed to promote quality, and the difficulty in achieving quality goals without support for modularity, scoping, and refactoring---a cost they had to bear to optimize exploration.
\mypara{Studies with Humans}
While corpus studies derive notebook usage patterns through notebook artifacts, interviews, observational studies, and surveys seek the answer directly from notebook users (mainly data scientists). 
\del{Through interviews and surveys, Kery et al.
found that data scientists prioritized exploration over explanation, which often backfire when one revisited the work at a later time.}
%
%
%
Wang et al.~\cite{wangHowDataScientists2019} conducted the first observational study where pairs of data scientists collaborated in a notebook task, focusing on collaboration patterns but not the low-level notebook-related actions.

\del{Two}\ins{Four} prior works\ins{, although targeting general data workers as opposed to scientsts,} are particularly relevant to ours.
\ins{Through interviews and surveys, Kery et al.}~\cite{kery2017variolite}\ins{ found that data scientists prioritized exploration over explanation, and this prioritization often backfired when revisiting work later; in contrast, we found that scientists’ priorities depend on context and goals. 
For example, P17 carefully documented exploratory code with Markdown notes, expecting to revisit his code years later, even though documentation could inhibit exploration later by increasing modification costs.
}
\ins{As }Subramanian et al.~\cite{subramanianCasualNotebooksRigid2020} \ins{found from}\del{analyzed} screen recordings of \del{21}\ins{nine} data workers performing their own tasks\del{ in their preferred setup---nine used Jupyter notebooks}, \del{and their recordings showed that notebooks became disorganized easily.}\ins{our study showed that scientists used notebooks for both experimentation and results sharing;}\del{Our study revealed quality attributes besides clarity that scientists valued.}\ins{ additionally, while this work showed that notebooks could easily become disorganized, most scientists in our study valued clarity in notebooks.}\del{ and enforced tactics to promote it, which could inhibit other quality attributes they valued (\eg, avoiding debugging within function definitions, which maintained clarity but attenuated debuggability).}
\ins{Our participants adopted tactics to promote clarity in their notebooks, but these same tactics often inhibited other quality attributes they valued (\eg, avoiding debugging within function definitions, which maintained clarity but limited debuggability).}
%
%
Chattopadhyay et al.
~\cite{Chattopadhyay2020:Whats} 
conducted observations, interviews, and surveys with industrial data scientists and engineers, revealing \ins{nine }pain points \del{in}\ins{when} using notebooks\del{ consistent with our eight software quality attributes that scientists valued in notebook work, but the study population differs from ours. In addition, we focus on the design decisions of notebooks that affected the pursuit of quality software and identified user tactics that promote quality.}\ins{. Although their study population differs from ours, our participants encountered similar issues, including a lack of built-in (AI) code assistance tools and difficulty refactoring code, hitting all pain points except data security; in addition, our work surfaced tactics scientists adopted to work around the pain points (\eg, P20 split a notebook into two sections to address limitations in collaborative work).}
\ins{Finally, in addition to conducting two corpus studies, Rule et al.}~\cite{Rule2018:Exploration}\ins{ interviewed 15 academic data analysts who felt that although messes built up easily, notebook clarity was unnecessary unless for sharing; in contrast, we found clarity to be a top quality attribute that scientists cared about, even without sharing or collaboration, with four participants explicitly using tactics to promote notebook clarity in non-collaborative work, such as adding documentation and writing modular code.
}%

\mypara{Notebook Improvements and Novel Systems}
Driven by the existing challenges with Jupyter notebooks, researchers have reviewed the design of computational notebooks~\cite{Lau2020:Design} and proposed analysis techniques~\cite{titovReSplitImprovingStructure2022, macke2021fine}, novel notebook systems~\cite{Observable, Horowitz2023:Engraft, natto}, and Jupyter extensions~\cite{keryStoryNotebookExploratory2018, Head2019:Managing, koop2017dataflow, macke2021fine, Wang2022:StickyLand, weinmanForkItSupporting2021} to improve the notebook quality and user experience.
Some systems incorporate informal version control into notebooks to help users explore and compare code alternatives~\cite{keryStoryNotebookExploratory2018, Wang2022:StickyLand, weinmanForkItSupporting2021}.
%
Some tackle the error-prone manual state management issue in Jupyter by reporting unsafe notebook executions that lead to out-of-sync data dependencies~\cite{macke2021fine} or allowing dataflow execution across cells~\cite{Observable, koop2017dataflow, natto}.
Other systems aim to reduce the potential clutter created during the exploration process by providing more live feedback~\cite{Horowitz2023:Engraft} and cleaning up redundant code with program slicing~\cite{Head2019:Managing}.
%
Our study complements these systems by proposing design opportunities for future improvements to Jupyter notebooks grounded in observational data.

\section{Conclusion}
Scientists value eight different quality attributes in their work and use 18 tactics to  promote those quality attributes. The cell model in computational notebooks promotes key quality attributes, such as \emph{explorability}, that scientists value, partly explaining their dominance among scientists. Although the model also inhibits other valued quality attributes, such as reproducibility, future changes to the notebook model could enable scientists to meet their quality goals and navigate the spectrum from \emph{exploration} to \emph{explanation} more effectively.

\section*{Acknowledgments}
We thank the anonymous reviewers for their suggestions.
This work was supported in part by NSF grant 2107397.

\end{sloppypar}

\newpage
\bibliographystyle{IEEEtran}
\balance
\bibliography{citations}

\end{document}